\documentclass[12pt]{article}
\usepackage{graphicx}
\usepackage{subfigure}
\newcommand{\be}{\begin{equation}}
\newcommand{\ee}{\end{equation}}

\begin{document}
\begin{titlepage}

\vskip2truecm

\begin{center}
\begin{large}
{\bf Slow dynamics in the 3--D gonihedric model}

\end{large}
\vskip1truecm

{\sc{P.~Dimopoulos} \footnote{E-mail: petros@ecm.ub.es },
D. Espriu \footnote{E-mail: espriu@ecm.ub.es },
E. Jan\'e \footnote{E-mail: ejane@ecm.ub.es },
A. Prats \footnote{E-mail:
prats@ecm.ub.es}}\\
Departament d'Estructura i Constituents
de la Materia \\ 
Universitat de Barcelona \\
Diagonal 647, 08028 Barcelona, Spain
\\
\end{center}
\vskip1truecm

\begin{abstract}
\noindent We study dynamical aspects of three--dimensional 
gonihedric spins by using Monte--Carlo methods. 
The interest of this family of models (parametrized by one
self-avoidance parameter $\kappa$) lies in their capability to show
 remarkably slow dynamics and seemingly glassy
behaviour below a certain temperature $T_g$  without the need
of introducing disorder of any kind.
We consider first a hamiltonian that takes into account only 
a four--spin term ($\kappa=0$), where a first order
phase transition is well established.
By studying the relaxation properties at low temperatures we confirm 
that the model exhibits two distinct regimes. For $T_g< T < T_c$,
with long lived metastability and a supercooled phase, the approach
to equilibrium is well described by a stretched exponential. For
$T<T_g$ the dynamics appears to be logarithmic. We provide
an  accurate determination of $T_g$.  
We also determine the evolution of particularly long lived 
configurations.
Next, we consider the case $\kappa=1$, where the 
plaquette term is absent and the gonihedric action consists
in a ferromagnetic Ising  with fine-tuned
next-to-nearest neighbour interactions. This model exhibits a 
second order phase transition. The consideration
of the relaxation time for configurations in the cold phase reveals 
the presence of slow 
dynamics and glassy behaviour for any $T< T_c$. 
Type II aging features are exhibited by this model.

\end{abstract}

\vspace*{6cm}
\noindent UB--ECM--PF 02/08 \\
\noindent April 2002
\end{titlepage}

\section{Introduction}
Glassy systems are very common in  nature, yet  not quite well understood.
Lattice models may serve as good candidates to describe some
properties of these systems.
In recent years some interest has been raised by remarkably simple
Ising spin systems that originated from
the consideration of a model of  random surfaces in the context of
string theory \cite{Amba-sukiasian}, \cite{savvidy}.
The version of this model in a discretized space defines the so-called gonihedric spin model
which consists, in general, of an Ising model with finely tuned
nearest, next-to-nearest neighbour and plaquette
interactions. The relation among the couplings of the hamiltonian depends also
on the dimensionality of the system. The geometric origins of the
model show up in a remarkable simple way of writing the energy of a
given configuration: the surfaces corresponding to the interfaces
between up and down spins are weighed by $E=n_2 + \kappa n_4$, where 
$n_2$ is the number of edges of such an interface and $n_4$ is the
number of four plaquettes that share a common link. The parameter 
$\kappa$ can thus be interpreted as an indicator of the
 self-avoidance of the model. Notice that there is
no microscopic surface tension.

Up to now there has been a considerable amount of
numerical work on the three--dimensional case, to which we shall refer in the
following and also some peliminiary results in four--dimensional case
\cite{kuts}.
In two dimensions, the model with $\kappa=0$ is actually trivial (no
phase transition \cite{bathas}) but the solution with $\kappa\neq 0$ 
is unknown.

For the three dimensional case which is of our concern in this paper the Hamiltonian
of the model takes the form
\begin{equation} \label{ham2}
H({\sigma})=-2\kappa \sum_{\vec{r},\vec{\alpha}} \sigma_{\vec{r}} \sigma_{\vec{r}+\vec{\alpha}} +
\frac{\kappa}{2}\sum_{\vec{r},\vec{\alpha},\vec{\beta}} \sigma_{\vec{r}}\sigma_{\vec{r}+{\vec{\alpha}}+
{\vec{\beta}}}-\frac{1-\kappa}{2}\sum_{\vec{r},\vec{\alpha},\vec{\beta}}
\sigma_{\vec{r}} \sigma_{\vec{r}+\vec{\alpha}}\sigma_{\vec{r}+\vec{\alpha}+
\vec{\beta}}\sigma_{\vec{r}+\vec{\beta}},
\end{equation}
where $\vec{\alpha}$ and $\vec{\beta}$ are lattice unit vectors. The
model is defined on an cubic lattice.
The system exhibits a very high degree of symmetry due to the particular ratio of the
couplings. This symmetry implies that there is no  cost
by the flipping of any plane
of spins. This results in
a highly degenerate ground state. Note in passing that this
last feature is common in every glassy system.

Some numerical evidence regarding dynamical and equilibrium properties
of the hamiltonian (\ref{ham2}) has been accumulated in the past, particularly
for $\kappa=0$, but also for $\kappa\neq 0$.  These studies revealed many interesting
features and provided evidence that a glassy phase is present in the
phase diagram \footnote{For the interesting case of the four--spin
model where randomly distributed  couplings are considered,
see (\cite{ritort}).}. First of all, it is known that the model has a first order phase transition
at  a temperature $T_{c}$ where the solid to liquid transition is
present \cite{esbaijo}. There is also good evidence that a
dynamical transition exists at  $T_{g}<T_{c}$ which 
seems to mark the onset of the glassy behaviour \cite{lip}.
The existence of $T_{g}$ and the study of the dynamical properties of the system above and
below that value have been considered
in \cite{esbaijo}-\cite{lipjo3}. Preliminary results for
$\kappa\neq 0$ were given in \cite{baig} in what respects the equilibrium
properties, and in \cite{eslipjo} in what respects the dynamical properties
of the system .

In section 2 we provide a qualitative view of the slow dynamics
behaviour of the cold phase by considering the
relaxation properties of the model and we estimate with good
accuracy $T_{g}$ by measuring the
spin-spin autocorrelation functions.
We show that there is a dramatic change in the
behaviour of this function above and below  $T_{g}$.
In section 3 we study a different case of the spin gonihedric action
by taking $\kappa=1$ in (\ref{ham2}), apparently a much simpler system (the
plaquette term in the hamiltonian is absent for this value).
For that value of the parameter $\kappa$ the system is described by
nearest and next-to-nearest interactions. We confirm, by studying the energy susceptibility,
that the system has a second order phase transition \footnote{Note that from
some preliminiary results referred in (\cite{esbaijo}) the first order
transition present for $\kappa=0$  gets
weaker and possibly becomes second order at $\kappa \sim 0.5$.} . Furthermore we find slow
dynamics behaviour anywhere in the
cold phase (below the critical temperature).
By studying  spin--spin autocorrelation function and the overlap spin
distribution function, we provide  evidence that the system
exhibits type II aging \cite{mezard}, which is a feature of the glassy systems.

The physical interest of this model is twofold. On the formal
side, the model is of such simplicity that a theoretical understanding
of the mechanisms underlying slow dynamics and
glassy behaviour appears possible. On a more practical side, it would be extremely
interesting to be able to understand and produce magnetic
materials and coatings with such finely tuned (or approximately so) 
couplings. The extremely long relaxation times would make
them very robust agains thermal noise and fluctuations, yet encoding
information there would be as simple as in a normal 
magnetic material. This possibility has been suggested in \cite{sav}.
While the plaquette term seems hard to imitate in real materials,
the fact that many of the interesting features persist for $\kappa=1$
makes perhaps such possibility less remote.

\section{Four--spin interaction ($\kappa=0$)}

\noindent In this section we will study the case $\kappa=0$, i.e, 
a spin model with only four--spin (plaquette)
interaction. In this case the hamiltonian (\ref{ham2}) takes the form:
\be \label{K0}
H=\frac{1}{2} \sum_{\vec{r},\vec{\alpha},\vec{\beta}}
\sigma_{\vec{r}} \sigma_{\vec{r}+\vec{\alpha}}\sigma_{\vec{r}+\vec{\alpha}+\vec{\beta}}\sigma_{\vec{r}+\vec{\beta}}
\end{equation}

This form of the interaction leads to a highly degenerated ground state.
Flipping every spin in any plane of the cubic lattice implies
invariant ground state energy. Taking this symmetry into account,
the degenerancy of the ground state is  equal to  $2^{3L}$ due to the $3L$ diferent
planes in a cubic lattice. This degeneracy survives even at $T\neq 0$.

\begin{figure}[!h]
\subfigure[]{\includegraphics[scale=0.30,angle=-90]{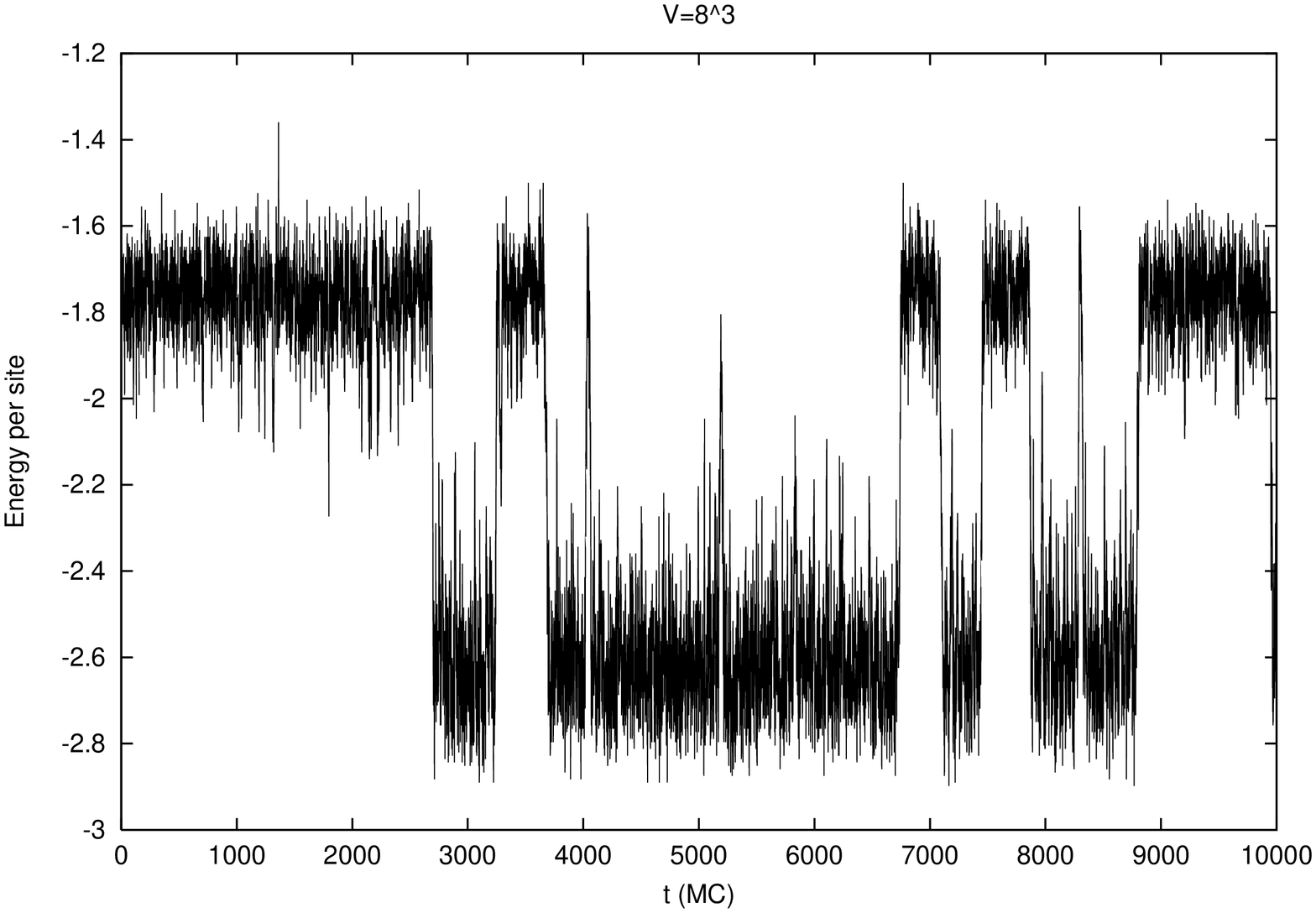}}
\subfigure[]{\includegraphics[scale=0.30, angle=-90]{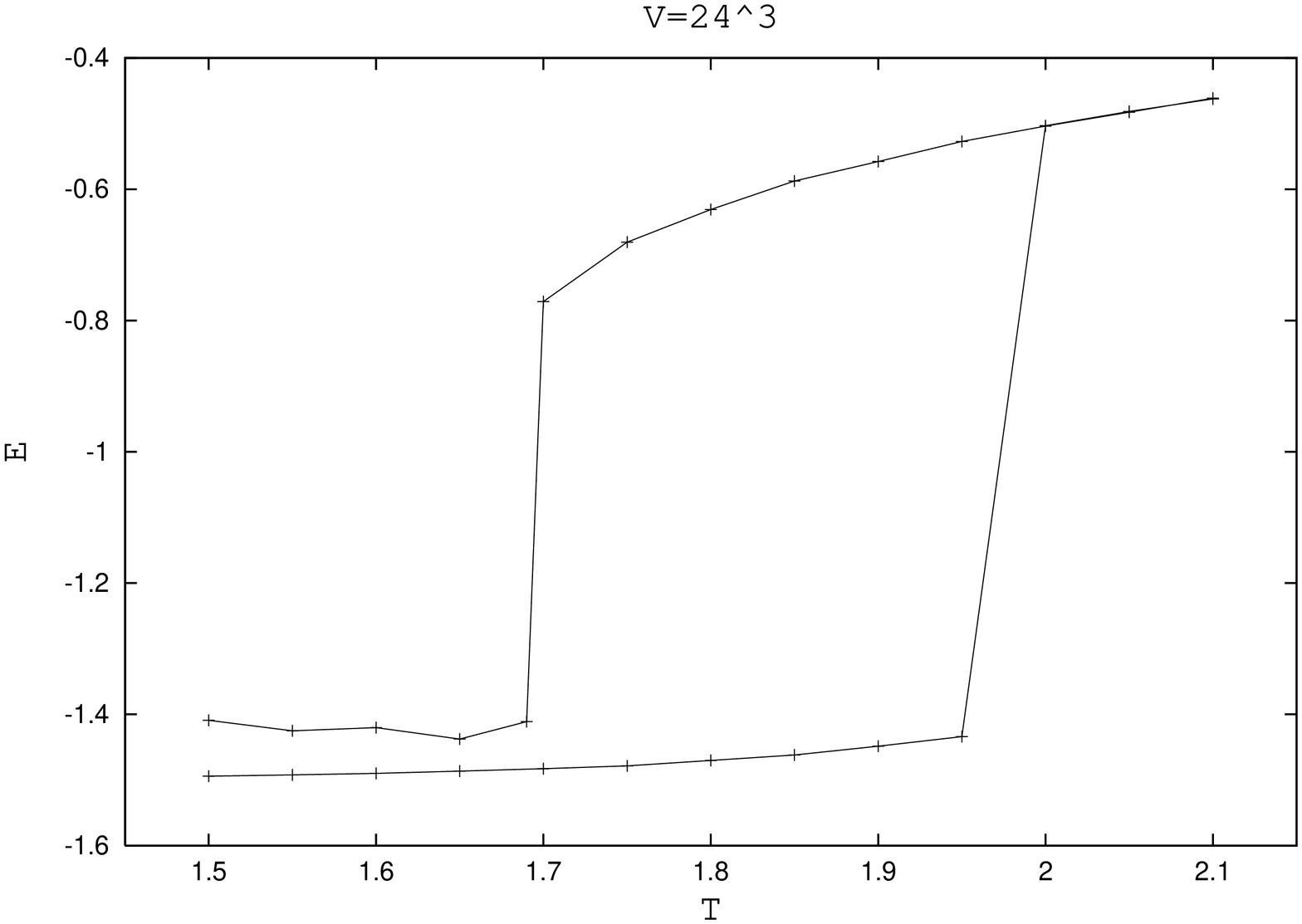}}
\caption{(a) An example showing the presence of metastability, (b) The jumping
from the supercooled to the equilibrium phase at $T=1.7$.}
\label{loop}
\end{figure}

It is well known  that this model exhibits a first order phase transition
at $T_{c}=1.95$ along with a dynamical transition at  $T_{g}\sim 1.7$, which is the temperature
where the  glassy behaviour shows up \cite{lip}, \cite{lipjo}, \cite{swift}.
Even though our main interest is the study of the glassy characteristics
by looking at  the relaxation as well as the autocorrelation 
 of the order parameters
(to be defined below) in the glassy phase, the region $T_g<T<T_c$ 
is interesting as well. In this region, numerical simulations  clearly 
indicate the presence of metastability. This is 
exemplified by the result presented in Fig.  \ref{loop}a where the time evolution of
an $8^3$  volume  at a temperature value just below  $T_{c}$ is shown \footnote{This simulation
refers to the system expressed in dual variables \cite{sav2} and it has been carried out 
using a cluster algorithm.}.  
This figure also gives clear evidence that there is a first order phase transition.
Fig. \ref{loop}b consists of two different curves. One of them corresponds to a heating process 
starting from an initial ordered congiguration. The other one describes the result of quenching a random
initial configuration for each one of the temperature values shown. In both cases the simulation  has been carried out 
using the Metropolis algorithm on a $24^3$ volume by performing $10^4$ measurements
at each temperature value. As we will see in a while, the approach to
equilibrium  in the region  $T_{g}<T<T_{c}$ is non-standard and it is well described
by a stretched exponential, instead of a simple exponential.
For $T<T_g$
one immediately sees that the results from the quench of the random configuration after $10^4$ MC steps 
differ from those obtained starting
from an ordered configuration (any of the $2^{3L}$ vacua) \cite{lip}. The difference
appears to be constant all over this region (for a fixed number
of thermalization steps). This clearly hints to the coexistence of two 
different dynamics. Initially fast dynamics quickly brings 
an initial configuration
which is badly out of equilibrium to some sort of approximate equilibrium.
At that point slow dynamics takes over and the evolution of the system is considerably freezed.

Before getting into the more quantitative aspects of these results, it
is perhaps interesting to turn to one of our motivations, namely to test
whether the appearence of slow dynamics makes the transition between 
two approximate ground states so slow as to make a given
configuration virtually indestructible by thermal fluctuations, thus
providing a convenient way of storing information.  

\begin{figure}[!h]
\subfigure[]{\includegraphics[scale=0.30,angle=-90]{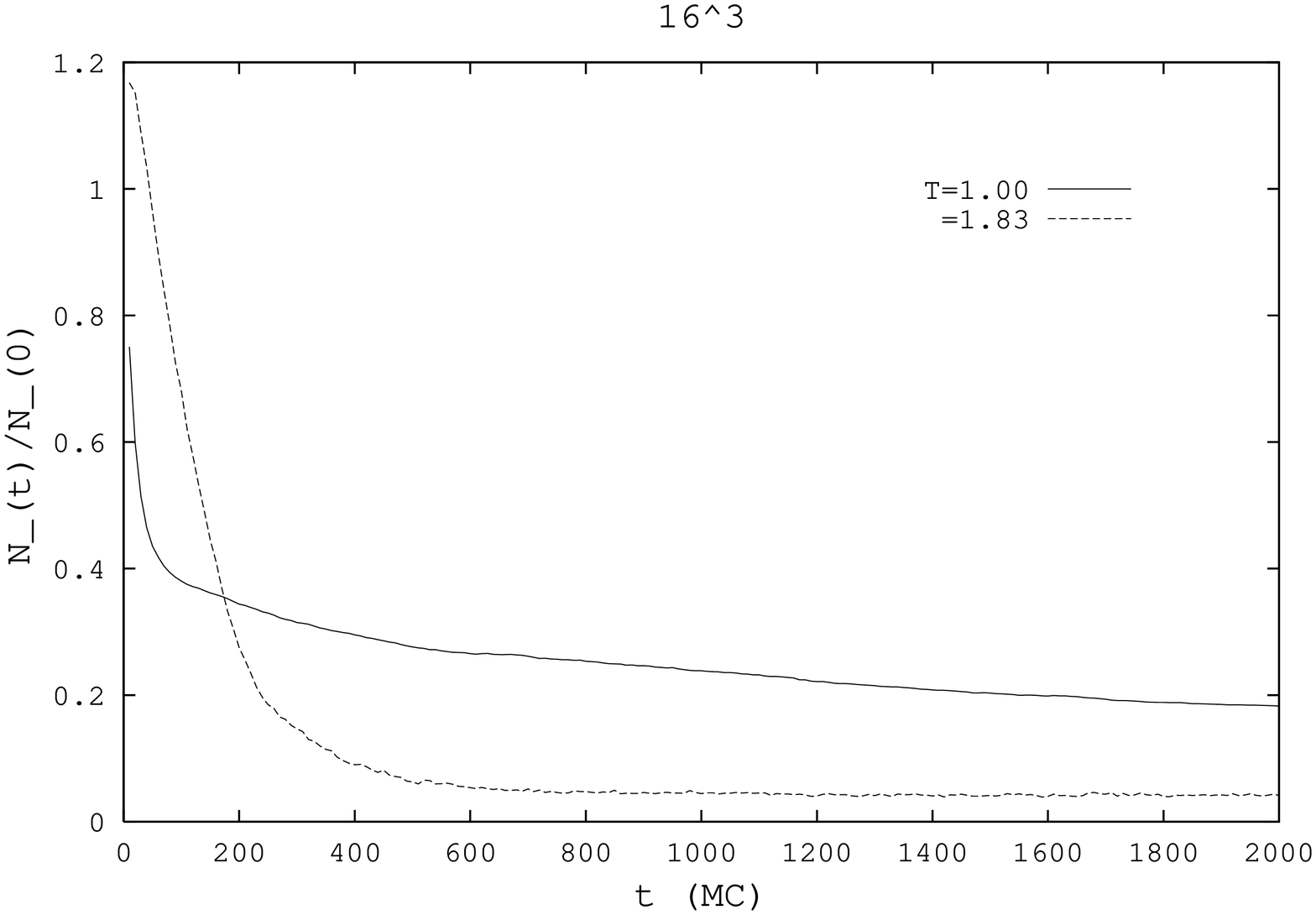}}
\subfigure[]{\includegraphics[scale=0.30, angle=-90]{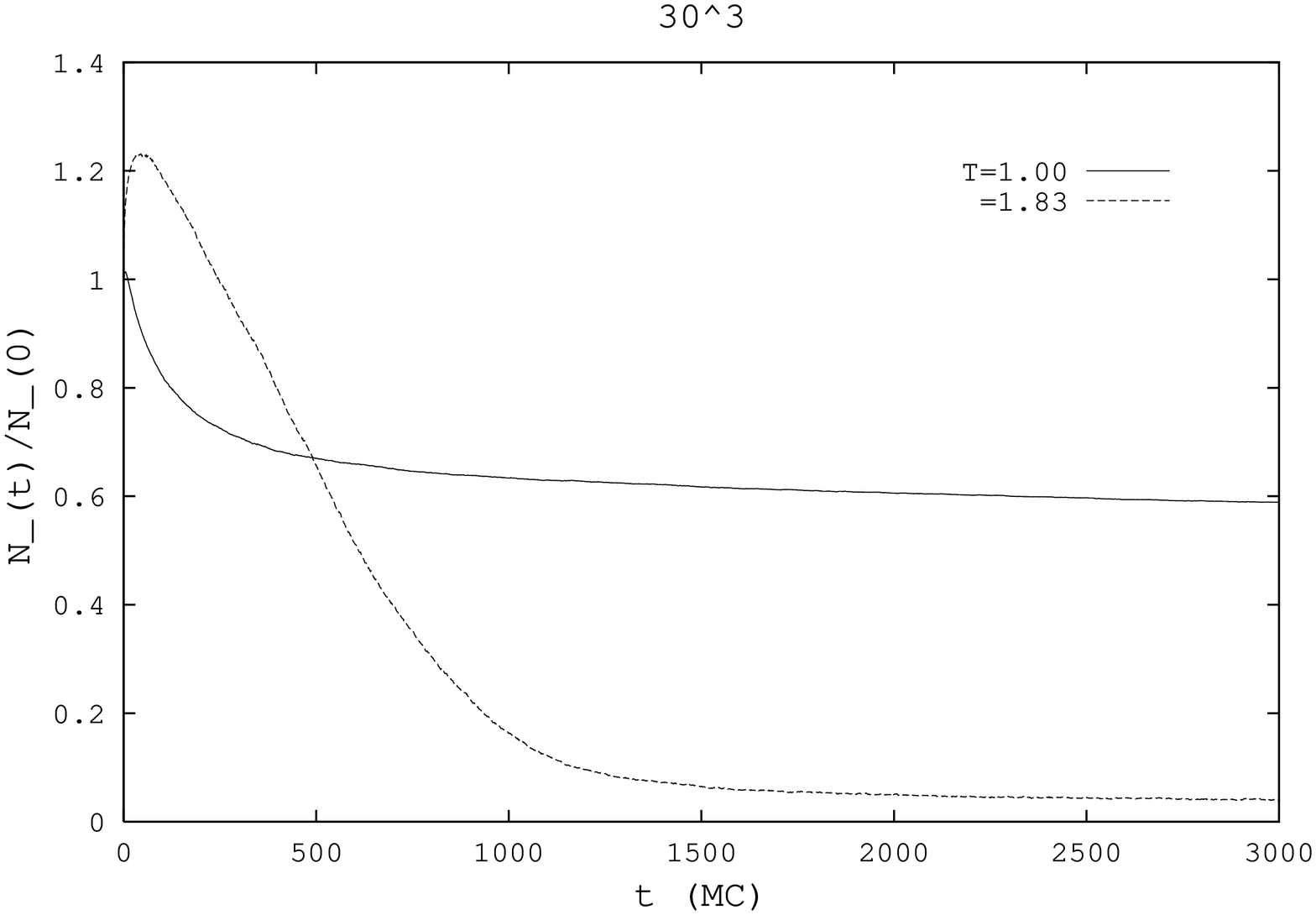}}
\caption{The relaxation for two diferent temperatures, $T=1.0$ and $T=1.83$:
(a) $16^3$ lattice volume with an initial chessboard--like configuration in a $8^3$ volume, (b) $30^3$ lattice volume
with an initial chessboard--like configuration in a $20^3$ volume.}
\label{fig1ab}
\end{figure}

To this end,
we simulate the system on a cubic lattice and we use a Glauber 
algorithm \cite{glauber}, which is assumed to provide a good approximation to the thermal
mechanism of fluctuations. Metropolis or heat bath give very similar
results. We look at the
decay of an artificial initial configuration consisted of an inner volume with a chessboard-like
arrangement of the spins, which is one of the ground states, while the outer volume spins
are fixed at $+1$ i.e, they form a ferromagnetic ground state.
We use fixed boundary conditions in order to make
the system decay to a ferromagnetic ground state.
Our measured quantity is the number of the minus spins, $N_{-}(t)$, in terms of the
computing time divided by  the  number of spins at time $t=0$, $N_{-}(0)$. The quantity
$N_{-}(t)$ is clearly related with the magnetization, $M$, since: $M \approx N_{+}+N_{-}=N_{tot}-2N_{-}$.
Two examples of our results are given in Fig. \ref{fig1ab}a  and Fig. \ref{fig1ab}b
which have been produced from simulations over
two lattice volumes namely $16^3$ and $30^3$ which at time $t=0$ have enclosed
$8^3$ and $20^3$ lattice volumes respectively with a chessboard--like arrangement for the spins.

\begin{figure}[!h]
\subfigure[]{\includegraphics[scale=0.30,angle=-90]{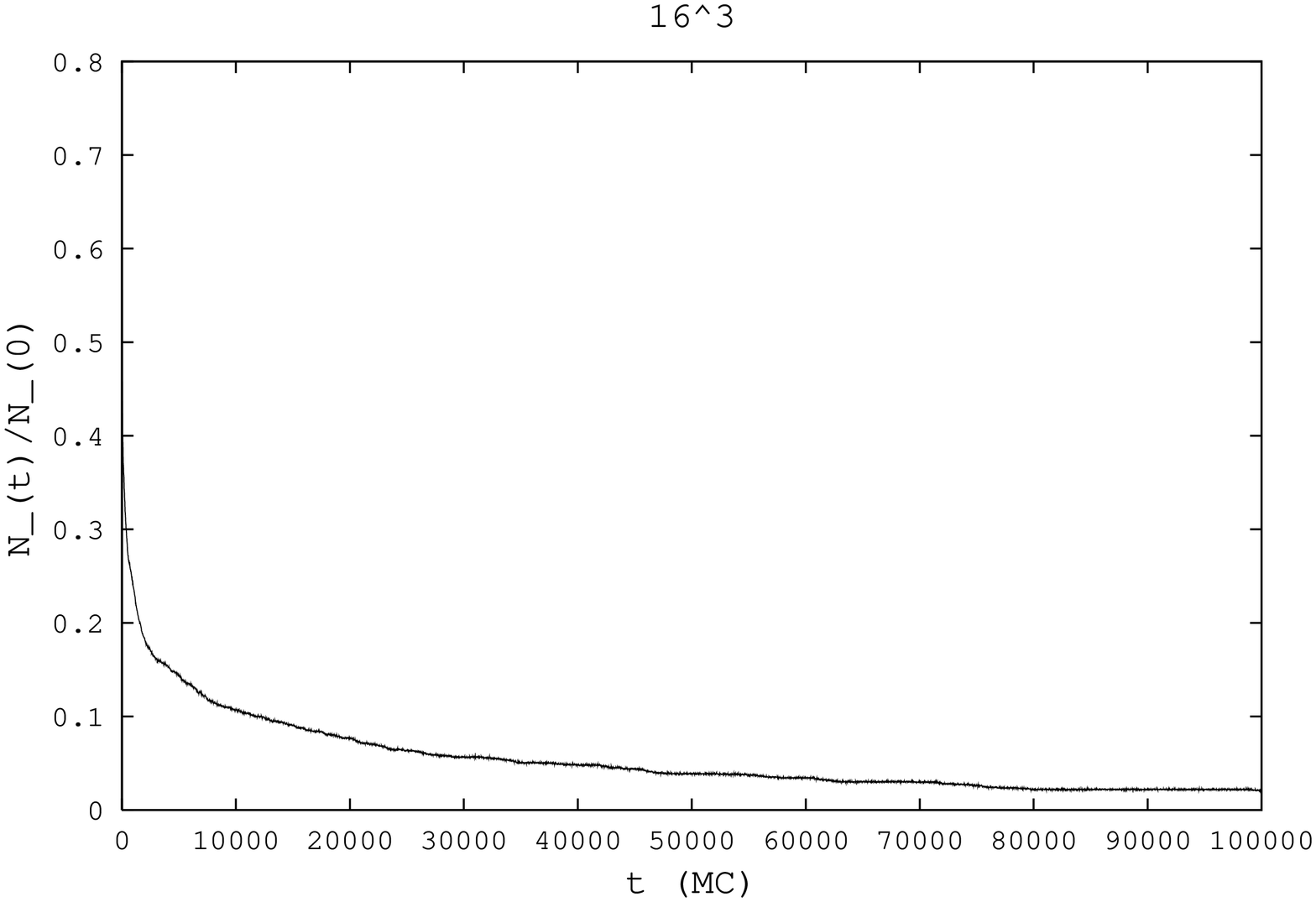}}
\subfigure[]{\includegraphics[scale=0.30, angle=-90]{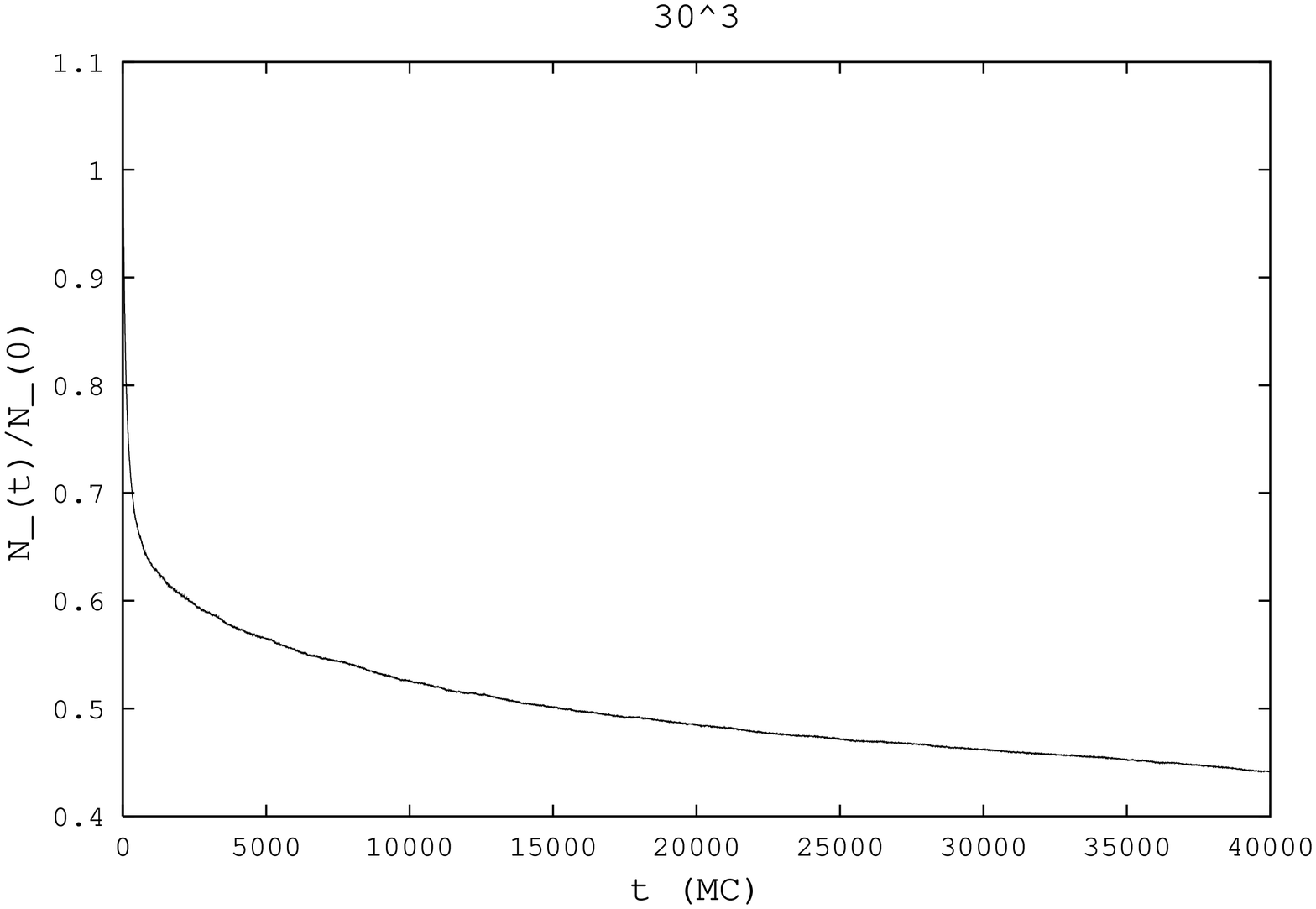}}
\caption{The long time  evolution at T=1.0 for inner chessboard--like configuration in a  $8^3$  (a) and
$20^3$ volume (b).}
\label{fig2}
\end{figure}

\begin{figure}[!h]
\begin{center}
\includegraphics[scale=0.40, angle=-90]{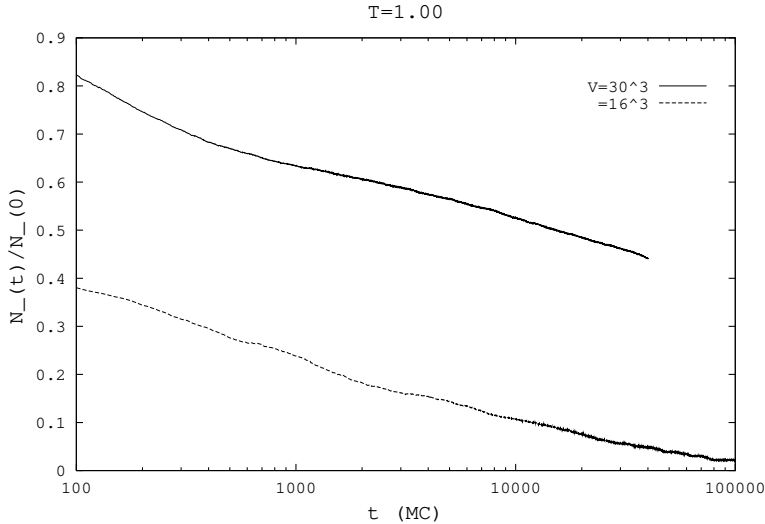}
\caption{The Fig. \ref{fig2} results as plotted in a logarithmic time scale. ($8^3$ and $20^3$ initial
chessboard-like configurations for total volumes $16^3$ and $30^3$ respectively.)}
\label{fig3}
\end{center}
\end{figure}

The two curves in each subfigure have been generated after averaging 
over a sample of 50--100 copies where
each one evolves starting from the
same initial configuration, in order
to reduce the noise. Two cases are shown. One is for temperature
$T=1.83$, in the
supercooled phase, while the other one, $T=1.0$, lies in the glassy phase .
The difference on the relaxation time between the two temperatures is obvious. For the
$T=1.83$ case  the system reaches a stable value relatively fast for both lattice volumes, but
this is not the case for $T=1.0$. Note that the slope of
the curve keeps 
taking non-zero value even for remarkably large times. This is perhaps
clearer in Fig. \ref{fig2} where long runs are depicted for $T=1.0$, 
confirming that we are in presence of slow dynamics.
This  behaviour can be seen in a more apparent way in Fig. \ref{fig3}
where the results of Fig. \ref{fig2}
are presented in a logarithmic time scale. In Fig. \ref{fig3} 
the logarithmic decay is present clearly enough.

\begin{figure}[!h]
\begin{center}
\includegraphics[scale=0.40, angle=-90]{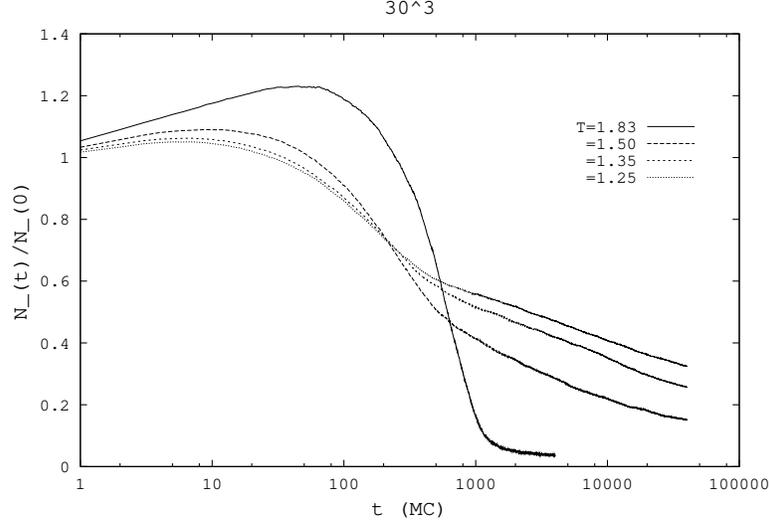}
\caption{For $T=1.83$ the system is in the supercooled region and exhibits fast relaxation. For the three 
other temperatures
values which stand in the cold phase the slow decay is obvious when it is presented in logarithmic time scale.}
\label{fig4}
\end{center}
\end{figure}

The very distinct dynamics between the glassy phase and the
evolution
of the supercooled phase in the metastability region $T_g<T<T_c$
can be seen in a more definite way in Fig. \ref{fig4}. In this plot the results for four
temperatures  are presented for the bigger volume used i.e. $30^3$.
All of them correspond to a random starting configuration.
For $T=1.83$, which lies in the metastability region the evolution seems very fast
all the way to  the  equilibrium value. On the contrary for the
other three temperatures, a fast
evolution is initially observed followed by a very slow one which
persists up to very long times.

These results are very suggestive and indeed show that the system
finds very difficult to overcome dynamical energy barriers that are
created along the evolution and this is undoubtedly the reason for the
slow dynamics. Recall that the energy of the model is concentrated
on the edges; the system has vanishing microscopic surface tension.
To reduce the volume of the excitation with local moves, the total
edge length must temporarily increase by a substantial amount.
That makes excitations such as the one we have been analysing
virtually stable.

In spite of the
unambiguity of the previous results, in order to determine 
some properties associated to the very slow dynamics observed at low
temperatures
we shall proceed to analysing several dynamical correlators.

\begin{figure}[!h]
\begin{center}
\includegraphics[scale=0.40, angle=-90]{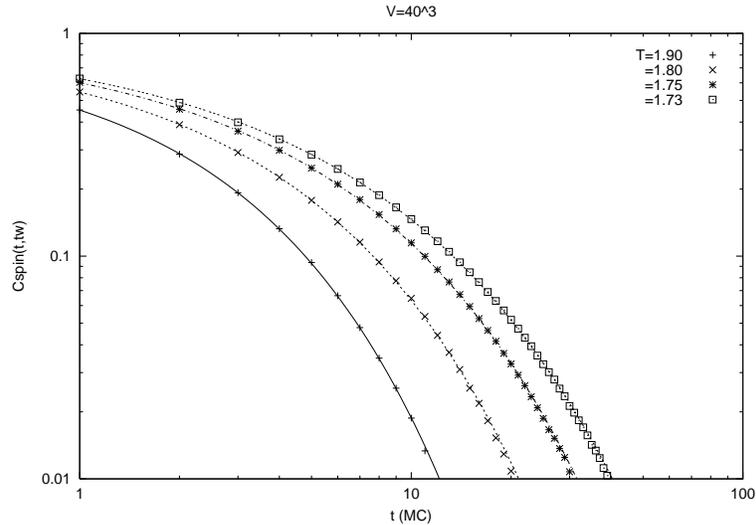}
\caption{Spin--spin autocorrelation function versus time. The fittings to the points are of
stretched exponential type.}
\label{CtK0}
\end{center}
\end{figure}

\begin{figure}[!h]
\begin{center}
\includegraphics[scale=0.40, angle=-90]{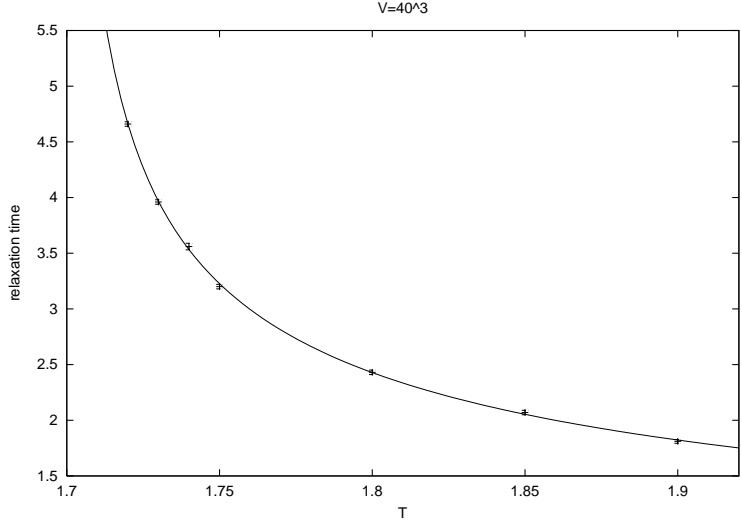}
\caption{$\tau$ versus $T$. The fitting procedure leads to divergent $\tau$ for $T_{g}=1.698$.}
\label{Ttau}
\end{center}
\end{figure}

We shall first study  the spin--spin autocorrelation function for 
temperatures lying in the metastability region after a random start,
thus forcing the system to be in a supercooled phase. Its
definition is given by
\be \label{cspin}
C_{spin}(t,t_{w})=\frac{1}{N} \langle  \sum_{\vec{r}} 
\sigma_{\vec{r}}(t_{w}) \sigma _{\vec{r}}(t+t_{w}) \rangle
\ee
The brackets mean that we take the average value over  copies starting from a random
configuration (200-400 in our case). The waiting time $t_{w}$ is the
time for which the system is
being thermalized before  taking  the measurements at subsequent times denoted by $t$.
The waiting time $t_{w}$ is taken to be about 300 by noticing that
from that value on 
the resultant values for $C_{spin}$  are quite identical as long as the
temperature lies in the metastability region. In Fig. \ref{CtK0} we
present some of our results for  $C_{spin}$ in a $40^3$ lattice volume
and for four temperature values. The fittings are  stretched exponentials of the form
$a e^{-(t/ \tau)^b}$. We denote by $\tau$ the relaxation time. 
In all cases we found $0.60<b<0.80$ with
an error smaller than 0.004. Since the fits seem quite good we go on
and make  a plot of the 
resulting  values of $\tau$ versus the temperature $T$. The result is
shown in Fig. \ref{Ttau} where the corresponding fit
to a function having the form, $\mbox{const}/(T-T_{g})^{c}$, is quite
good and leads to the prediction $T_{g}=1.698(1)$
with $c=0.41$.
At $T_{g}$ the autocorrelation time is expected to diverge because of
the onset of the slow dynamics which turns the stretched exponential
behaviour into a power law (with a small exponent) or a logarithm. 

Although that result is in good agreement with previous simulations
\cite{lip}, \cite{swift}
the method relying on the stretched exponential fit may prove to be
too risky for an exact prediction of $T_{g}$ due to
ambiguities in the fitting process and perhaps is not trustable to
that accuracy. As an alternative and a cross check, 
for two different temperature
values, namely $T=1.720$ and $T=1.695$ we show the behaviour of the
$C_{spin}$ for 
two very different values of the waiting time,
$t_{w}=300,2000$. One sees that for the higher temperature  the behaviour of
the $C_{spin}$ 
is identical but for the smaller one there is  a strong dependence on
$t_{w}$ 
showing that for larger waiting times the system exhibits
different, and in particular, slow dynamics behaviour which is a strong signal
that the system has passed to the glassy phase. Hence we are able to
estimate the value of $T_{g}$ to lie in between the interval $1.695 <
T_{g} < 1.720$ 
which indeed agrees also with the previously obtained value.

\begin{figure}[!h]
\begin{center}
\includegraphics[scale=0.40, angle=-90]{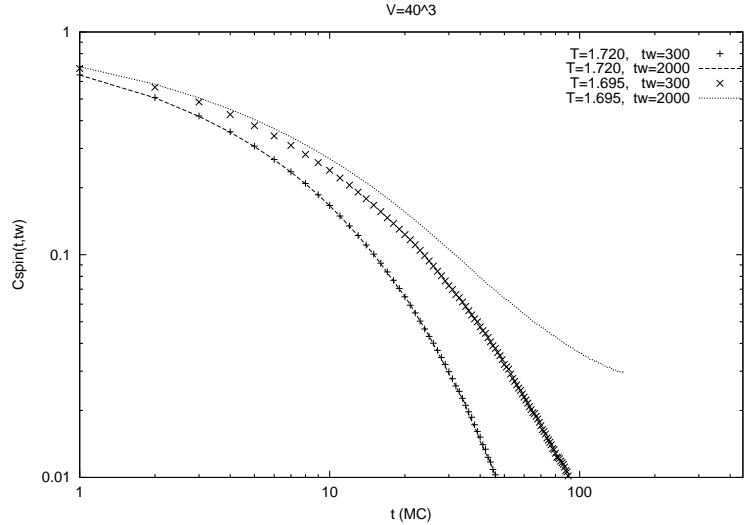}
\caption{$C_{spin}$ versus time for two values for the waiting time $t_{w}$. For the system being in the
glassy phase we see the dependence on $t_{w}$.}
\label{Ttaudif}
\end{center}
\end{figure}

\newpage
\section{Next and next-to-nearest neighbour interaction 
($\kappa=1$)}

\noindent
As we have indicated the gonihedric spin model is actually a family of
models which are parametrized by a real quantity $\kappa$.
All the members of this family share the common feature of having
their
spin interfaces weighed with the total edge length. The value of
$\kappa$ simply indicates the degree of self-avoidance of the surfaces.
 The plaquette
model whose dynamical properties we just discussed corresponds to
$\kappa=0$. An interesting member of this family is given by $\kappa=1$. The 
degeneracy of the ground state in this case is $3 \cdot 2^L$. 

In this case the Hamiltonian  is given by
\begin{equation} \label{ham}
H=-2 \sum_{\vec{r},\vec{\alpha}} \sigma_{\vec{r}} \sigma_{\vec{r}+\vec{\alpha}} + \frac{1}{2}
\sum_{\vec{r},\vec{\alpha},\vec{\beta}} \sigma_{\vec{r}}
\sigma_{\vec{r}+\vec{\alpha}+\vec{\beta}}
\end{equation}
\noindent As we see, the spin plaquette term has disappeared. From 
a practical point of view this model may be particularly interesting
as the plaquette term is obviously hard to get in real materials.  
From the standpoint of a spin system this is just a model
with nearest and next-to-nearest neighbour interactions (though a
finely tuned one).

We shall proceed to studying the dynamical properties of
the system by using Monte--Carlo methods.
In performing the simulations we used the Metropolis algorithm for several lattice
volumes, namely $10^3$, $16^3$, $20^3$, $24^3$, $30^3$, $40^3$ and
$46^3$. We imposed 
either fixed or periodic boundary conditions depending on the kind of
the measurement that 
was performed. During the presentation
of our results it will
be explicitly mentioned when we used one or the other.

\begin{figure}[!h]
\begin{center}
\includegraphics[scale=0.40, angle=-90]{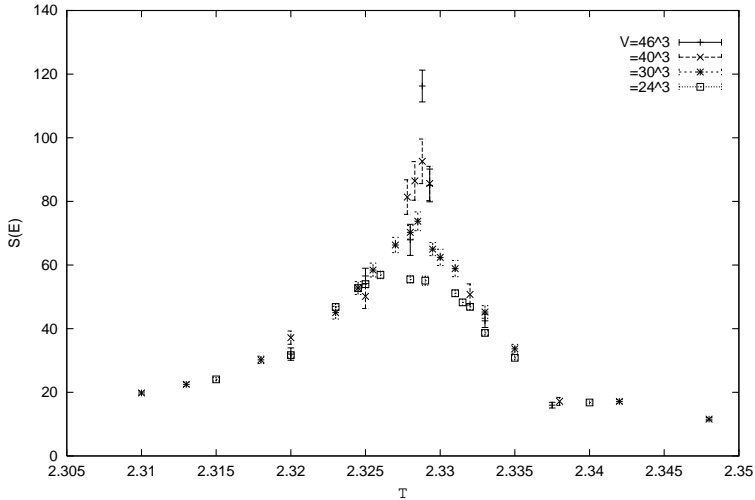}
\caption{$S(E)$ versus $T$ behaviour with the lattice volume.}
\label{fsusc}
\end{center}
\end{figure}

We begin by giving the behaviour for  the susceptibility of the energy
with the lattice 
volume which is given by
$$S(E)=V(\langle E^2\rangle -\langle E^2\rangle) $$
We denote by $V$ the lattice volume and by the symbol $\langle
\rangle$, the 
average value over sweeps.
In Fig. \ref{fsusc} the susceptibility for the system energy as a
function of the temperature for four lattice volumes is
depicted. For every point in the figure we performed $10^5$ thermalization sweeps followed by
more than $10^5$ measurements, using periodic boundary conditions and starting
from the ordered configuration.
The peak for each volume clearly increases with it although with an 
exponent less than one, which is a signal for
a second order phase transition. Also, the positions of the peaks for
the bigger volumes are seen to concetrate around
the value $T_{c}=2.329$ which is the pseudocritical value for the
up to the volumes we used. It should be mentioned that while the second
order character of the phase transition was already 
mentioned in  \cite{bathas}, \cite{jonmal} however our prediction for
$T_{c}$  gives a somewhat smaller value.

\begin{figure}
\subfigure[]{\includegraphics[scale=0.30,angle=-90]{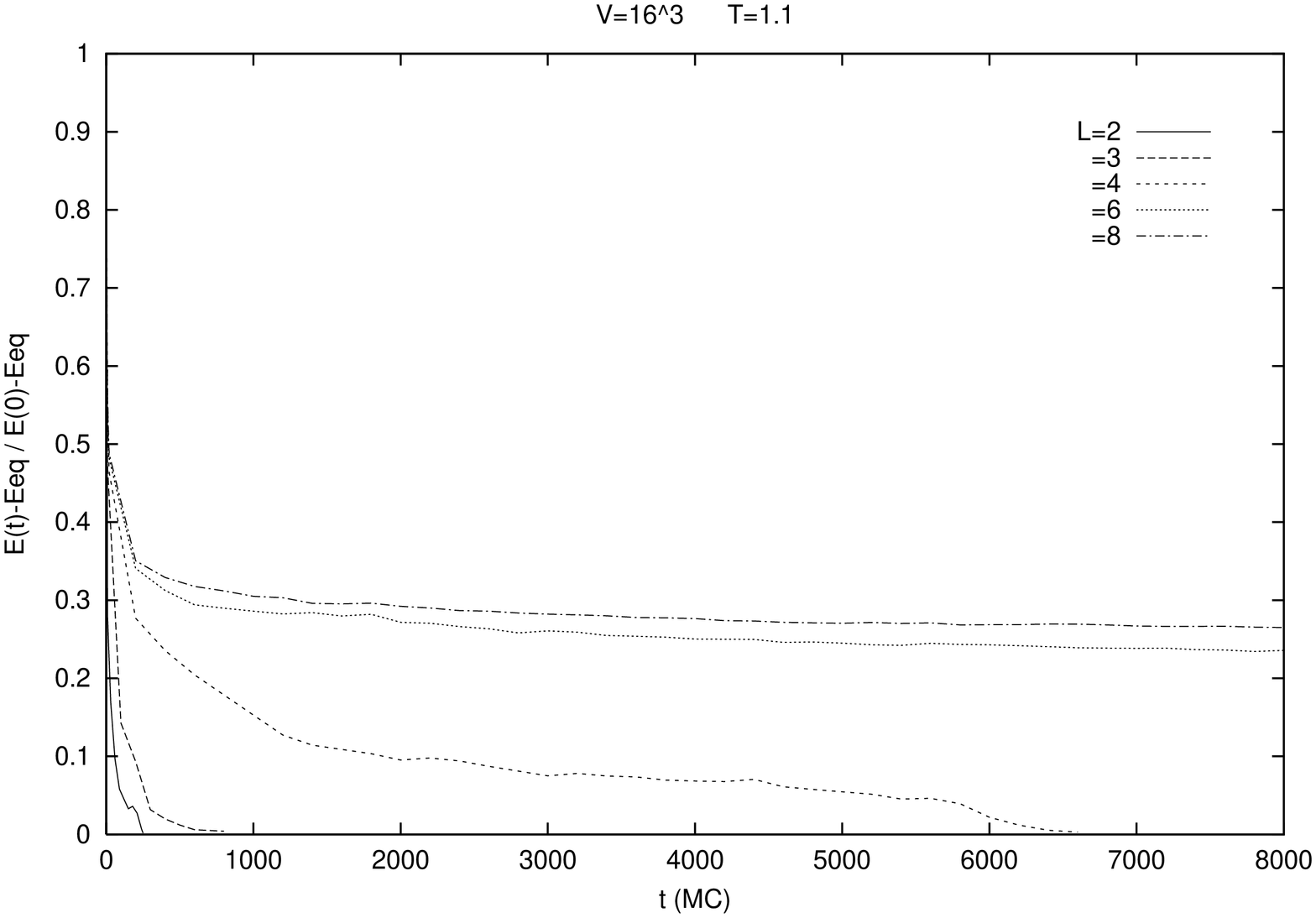}}
\subfigure[]{\includegraphics[scale=0.30, angle=-90]{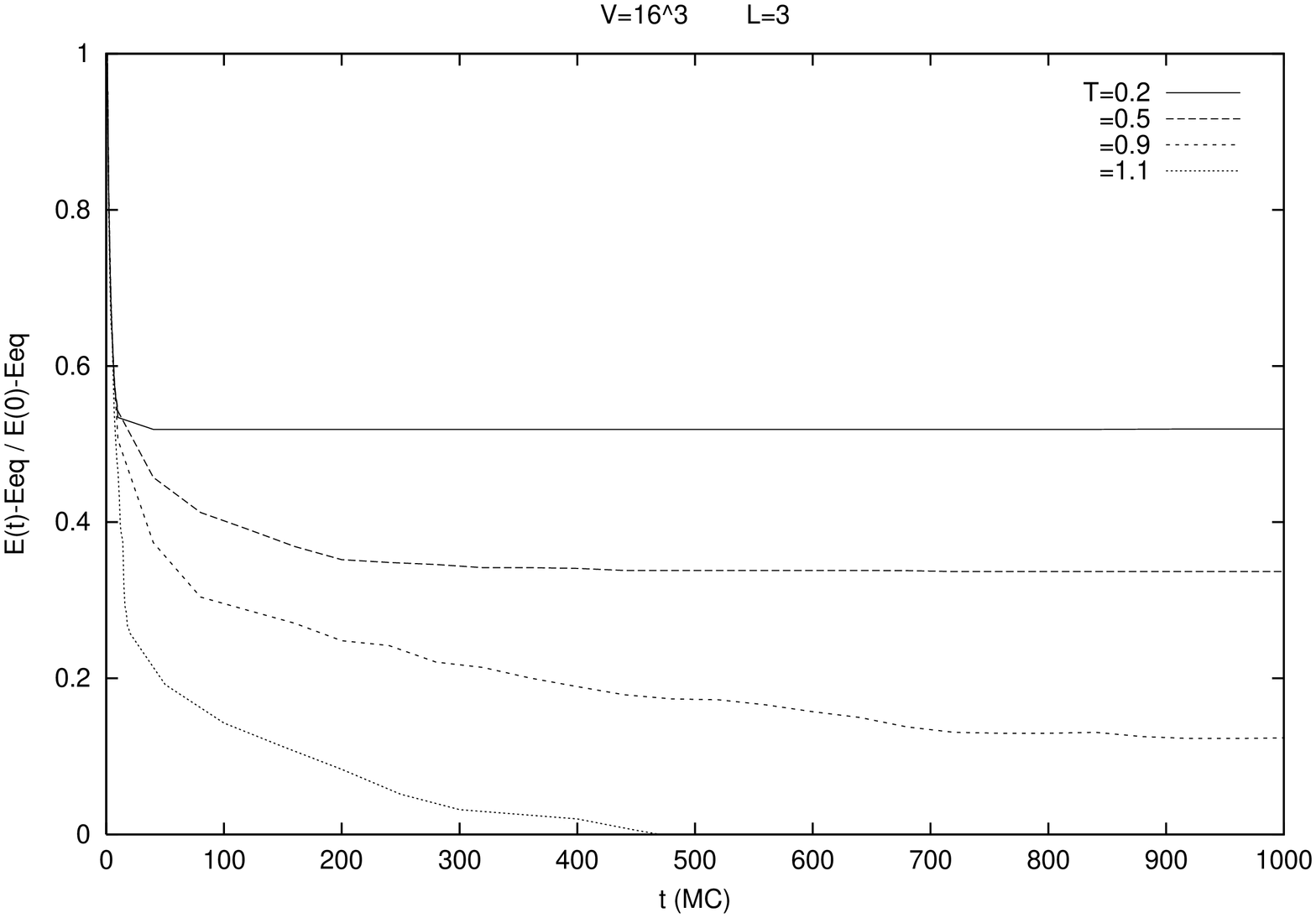}}
\caption{The relaxation behaviour of the initial configuration as seen by the study of quantity defined
in (\ref{Et}): a) for various inner volume sizes at fixed temperature (from top to bottom, $L=2,3,4,6,8$) and
b) for fixed inner volume size with decreasing temperature (from top to bottom $T=0.2,0.5,0.9,1.1$).}
\label{fv16}
\end{figure}

Having estimated the critical temperature value for the system we want 
now to study the dynamic behaviour 
at low temperatures, i.e. below $T_c$. To this end we use first the
same method as 
for the $\kappa=0$ case, described in the previous section, which
consists of choosing an initial chessboard--like configuration defined
in a cubic subvolume of the system.
We impose fixed boundary conditions and we study the behaviour of the
dynamical quantity
\be \label{Et}
\frac{E(t)-E_{eq}}{E(0)-E_{eq}}
\ee
with time for various temperatures in the cold phase and for different
sizes of the inner chessboard--like
volume. In Eq.(\ref{Et}) $E(t)$ and $E(0)$ denote the energy of the
system at time $t$ and just after the first
sweep performed, respectively and $E_{eq}$ is the system energy when an 
ordered configuration is taken as the initial one. 
In Fig. \ref{fv16}a we give an example for lattice volume $16^3$. We
fix the 
temperature at the value $T=1.1$ and we are interested in the
behaviour as the dimension of the inner chessboard--like volume
increases from $L=2$ to $L=8$. 
Each curve has been produced after
averaging several hundreds of repetitions starting from the same
initial configuration in order for the
noise to be reduced. The results show a dramatic
increase  of the relaxation time with increasing system size. 
For the maximum value of $L$ used, the
observable (\ref{Et}) shows very slow variation with the time
and it follows a power law behaviour with estimated
exponent equal to 0.18, so a logarithmic behaviour cannot be really
excluded.

We can reach similar conclusions by studying the above relaxation quantity
for decreasing values of temperature at fixed $L=3$.
In Fig. \ref{fv16}b the corresponding results are depicted for a
$16^3$ total volume.
In particular for the value $T=0.2$ the
dynamics is so slow that it seems really hard for the system to get to 
the ground state for accessible computing times.

Next we shall move to a more quantitative study 
following the same lines of the previous section. 
We look for aging features by considering the autocorrelation
spin--spin function defined by (\ref{cspin}).
\begin{figure}[!h]
\begin{center}
\includegraphics[scale=0.40, angle=-90]{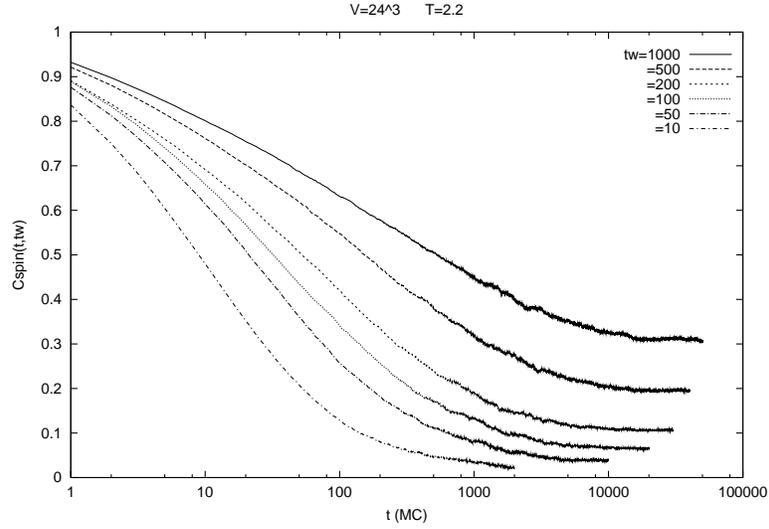}
\caption{The spin--spin autocorrelation function as depicted for different waiting times $t_{w}$
(from top to bottom $t_{w}=1000,500,200,100,50,10$).}
\label{C}
\end{center}
\end{figure}
Brackets indicate that we take the average value over 100-200 trials starting from a random
configuration. $t_{w}$ is the time for which the system is
being thermalized before  taking  the measurements at subsequent times
denoted by $t$. We follow the same lines
of analysis as in \cite{swift} which concerns the $\kappa=0$ case. In Fig. \ref{C} we show an example of
the behaviour  of the autocorrelation function (\ref{cspin}) for the
temperature value T=2.2 
and for a $24^3$ lattice volume. Six different curves for the
$C_{spin}$ are depicted each corresponding to a different value of $t_{w}$.
The very slow evolution of the autocorrelation function combined with the clear dependence on the
time $t_{w}$ give strong evidence in favour of aging. In order to
determine the aging 
(whether I or II) we consider
the overlap function \cite{mezard}:

\be \label{Q}
Q(t+t_{w},t+t_{w})= \frac{1}{N}\langle \sum_{\vec{r}} 
\sigma_{\vec{r}}^{(1)} (t+t_{w}) \sigma_{\vec{r}}^{(2)} (t+t_{w}) \rangle
\ee

\begin{figure} [!h]
\begin{center}
\includegraphics[scale=0.40, angle=-90]{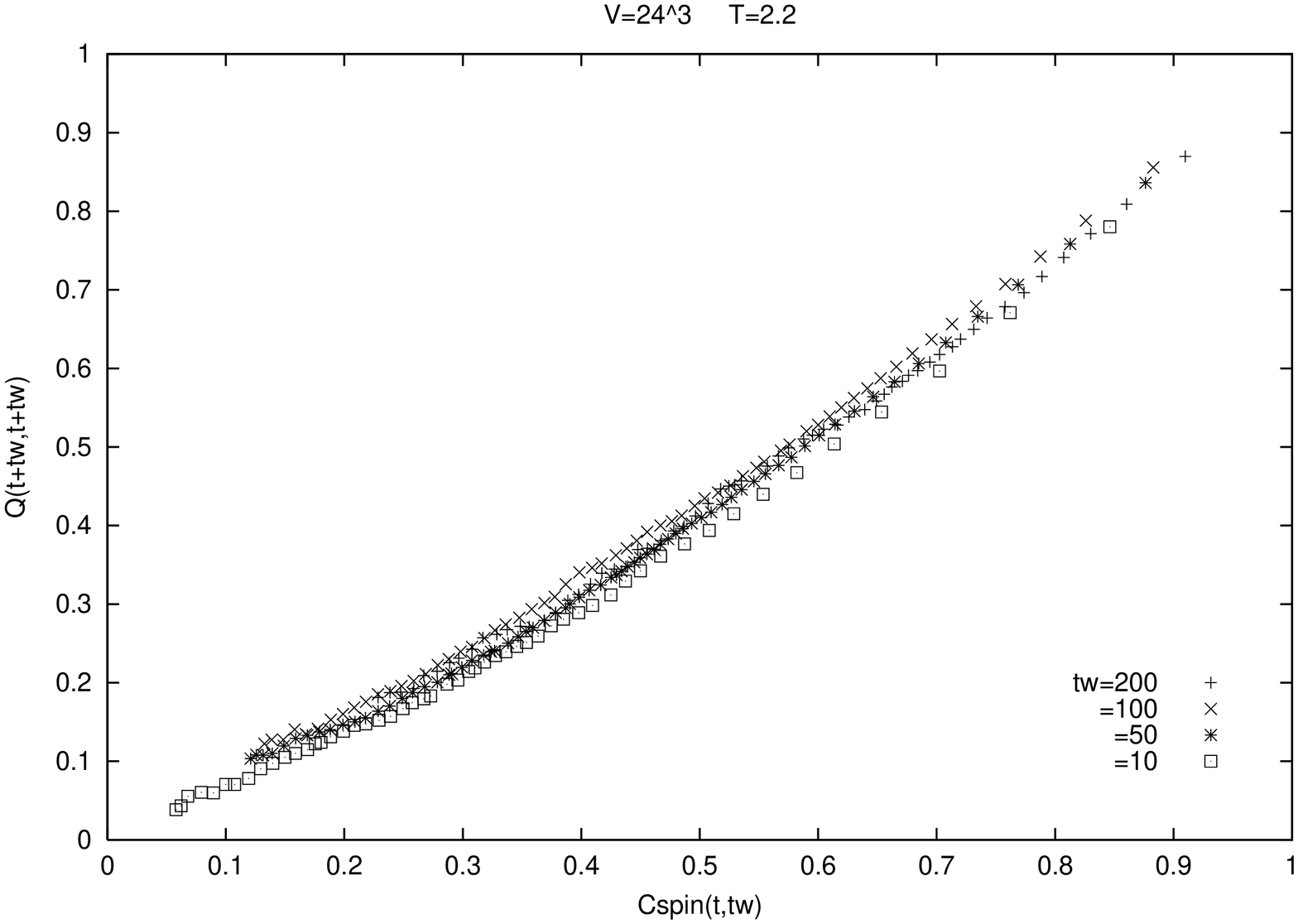}
\caption{The $Q(t+t_{w},t+t_{w}) \rightarrow 0$ behaviour as $C_{spin}(t,t_{w}) \rightarrow 0$ is clearly
seen for the four values of $t_{w}$.}
\label{QC}
\end{center}
\end{figure}

The measuring process for the above consists of the relaxation of the system for time $t_{w}$ starting from the
random configuration. At that moment we take two copies $\sigma^{(1)}$ and $\sigma^{(2)}$ each evolving
independently for time $t$. In Fig. \ref{QC} the behaviour of $Q(t+t_{w},t+t_{w})$ in terms of $C_{spin}(t,t_{w})$
for the same values for the volume and temperatute as in Fig. \ref{C} and for four values of the waiting time
$t_{w}$ is shown. It clearly follows that as $C_{spin}(t,t_{w}) \rightarrow 0$, $Q(t+t_{w},t+t_{w}) \rightarrow 0$. The
fact that the two system copies are moving independently since they show zero overlapping
indicates strong evidence for type II aging.

\section{Conclusions}

At this point we would like to summarize our main
conclusions.

We have carried out a rather complete analysis of the dynamical
properties of the gonihedric spin model. The model
naturally exhibits large potential barriers that are
dynamically generated along the evolution process when trying
to adjust to the environment. This is due to the fact
that the area term of the interfaces is
completely subleading and the dynamics is driven
by the total edge length of the excitations. This creates
basins of stability from where it is virtually
impossible to kick out the system by local
thermal fluctuations.

Several algorithms have been used: Glauber dynamics, heat bath
and standard Metropolis. Modulo an overall rescaling of Monte--Carlo
time, the results are fully equivalent.
  
In this  paper we have studied numerically two cases of 
the gonihedric spin model. For the $\kappa=0$ case  we give evidence of a
 glassy transition  and we estimate the temperature value $T_{g}$ at which the 
glassy phase arises being in the interval 
$[1.695, 1.720]$. We show that the dynamics is most likely
logarithmic for $T< T_g$. In the mestatability region, above
$T_g$ but below the true thermodynamical transition $T_c$ (where the
system has a first order transition) the 
approach to equilibrium is well described by a stretched exponential,
whose
characteristic time scale diverges as $T\to T_g^+ $.

We confirm that for the case $\kappa=1$ the system exhibits 
a second order phase transition. 
The critical properties of this transition have not been elucidated
yet;
a tentative determination of the exponents using finite size scaling
carried out in \cite{baig} gives a rather non-standard set of values.
Work on this is in progress, but the problem is agravated by
the presence of slow dynamics.
Indeed we 
have detected in the cold phase of this system very slow
dynamical behaviour, quite similar to what
takes place with $\kappa=0$ for $T< T_g$. This dynamical
behaviour has all the features of
type II aging and is 
again compatible with logarithmic evolution.

Possible applications of these simple, but interesting, spin
models are discussed.

\noindent {\bf Acknowledgements}
\vspace*{0.3cm}

\noindent D.E. thanks D. Johnston and A. Lipowski for discussions and the 
hospitality of the Department of Mathematics of Heriot--Watt University.
P.D. is grateful to G. Koutsoumbas for useful discussions and acknowledges support from "EUROGRID--
Discrete random geometries: from solid state physics to quantum gravity" (HPRN--CT--1999--00161).
The support from grant MCyT FPA 2001--3598 and CIRIT 200156R--00065 is also acknowledged. Some of 
this work was carried out at CESCA.

\end{document}